\newcommand{\Tr}{\mathop{\mathrm{Tr}}\nolimits}

\documentclass[osajnl,twocolumn,showpacs,eqsecnum]{revtex4}
\usepackage{amsfonts,amssymb,bm,graphicx}

\begin{document}

\title{Simple quantum  model for light depolarization}

\author{Andrei B. Klimov}
\affiliation{Departamento de F\'{\i}sica,
Universidad de Guadalajara, Revoluci\'on 1500,
44420~Guadalajara, Jalisco, Mexico}

\author{Jos\'e L. Romero}
\affiliation{Departamento de F\'{\i}sica,
Universidad de Guadalajara, Revoluci\'on 1500,
44420~Guadalajara, Jalisco, Mexico}

\author{Luis L. S\'anchez-Soto}
\affiliation{Departamento de \'{O}ptica,
Facultad de F\'{\i}sica, Universidad Complutense,
28040~Madrid, Spain}

\date{\today}

\begin{abstract}
Depolarization of quantum fields is handled
through a master equation of the Lindblad type.
The specific feature of the model is that
it couples dispersively the field modes
to a randomly distributed atomic bath.
The depolarizing dynamics emerging from
this approach is analyzed for relevant states.
\end{abstract}

%\ocis{270.0270, 260.5430, 270.2500}

\maketitle

\section{Introduction}

Polarization of light is a fascinating concept
that has deserved a lot of attention over the
years. Today, the field is witnessing a
renewed interest because of the fast developments
in optics, both on the applications and on the
more fundamental aspects. Moreover, the polarization
of a field is a robust characteristic, which is
relatively simple to manipulate without inducing
more than marginal losses. Therefore, it is not
surprising that many experiments at the
forefront of quantum optics use polarization
states~\cite{Asp81,Kwi95,Tse00,Bar03}.

The Stokes parameters provide, perhaps, the most
convenient description of polarization in classical
optics and can also be extended in a natural way
to the quantum domain~\cite{Jau59,Lui00}. They have
the drawback in that they measure only second-order
correlations of the field amplitudes: while this
may suffice for classical problems, higher-order
correlations are important for the realm of
quantum optics. This is the basic reason why the
Stokes parameters do not distinguish between
quantum states having quite different polarization
properties~\cite{Kly92,Usa01}.

Despite these subtleties, the Stokes parameters still
provide the most commonly used definition of a
degree of polarization~\cite{Bor99,Alo99,LWG03,KSY05},
as we discuss in Sec.~\ref{su2}. The degree of
polarization is often desired to reach its maximum
value, as well for classical as for quantum
communication. The term depolarization has come
to mean the decrease of the degree of polarization
of a light beam when traversing an optical system.
Our intuition strongly suggests that depolarization
is due to decoherence. Roughly speaking, two basic
mechanisms have been proposed to explain classically
this phenomenon~\cite{Bro98}: decorrelation of
the phases of the electric field vector and
selective absorption of polarization states.
The first cause is far more interesting to examine
when the microscopic mechanisms responsible of
depolarization are analyzed.

To deal with this decoherence, some ingenious
and practical tricks have been proposed, such as
coupling each mode to a beam splitter where it is
mixed with a vacuum~\cite{Dur04}. We wish to look
at this problem from a more fundamental point of
view. Of course, one can find in the literature
a number of pure dephasing models in which the
environment is represented as a scattering
process, which does not lead to a change of
populations~\cite{CL81,LCD87,Gar00,Yu02,SS04}.
For many applications, a good knowledge of
dephasing is of utmost importance. This holds
most prominently for quantum information
processing, where the operation completely
relies on the presence of coherence. It is
hardly surprising that a lot of attention
has been paid to dephasing in systems such as
quantum dots~\cite{Tak99,UJT00,KAK02,Paz02},
Josephson junctions~\cite{MSS99,NPT99,WHW00},
or general quantum registers~\cite{RQJ02}, to
cite only a few relevant examples.

However, the study of depolarization is not only
a question of pure dephasing dynamics. Polarization
has an additional su(2) invariance that leads to
a natural structure of invariant subspaces. On
physical grounds, we argue that this structure
must be preserved in the evolution, which
makes previous models fail in this case.
The main goal of this paper is precisely to provide
a simple approach that goes around this drawback
and provides a picture of the mechanisms involved in the
depolarization processes. Since in classical optics
the randomization is produced as light propagates
in a material medium, our main idea is to couple
the field modes dispersively to a randomly distributed
atomic bath: the resulting master equation has a
quite appealing structure that is examined in
some detail in Sec.~\ref{smf} for the case of
single-mode fields. Finally, in Sec.~\ref{mmf} we
extend the formalism to multimode fields, while
our conclusions are summarized in Sec.~\ref{con}.

\section{Polarization and su(2) invariance}

\label{su2}

In the interest of retaining as much clarity as
possible, we first recall some well-known facts
about the polarization structure of quantum fields.
Since the formalism applies to fields of arbitrary
wavefronts and frequencies, we consider a free
transverse electromagnetic field with $m$
spatiotemporal modes, whose positive-frequency
part is
\begin{equation}
\label{Epol}
\mathbf{E}^{(+)} (\mathbf{r}, t) = i
\sum_{j=1}^m \sum_{s= \pm}
\sqrt{\frac{\hbar \omega_j}
{2 \epsilon_0}} \ a_{js} \
\mathbf{u}_{js} (\mathbf{r})
\exp(- i \omega_j t) .
\end{equation}
In the Dirac quantization scheme the modes are plane
waves
\begin{equation}
\mathbf{u}_{js} ( \mathbf{r} ) =
\frac{1}{\sqrt{V}} \
\mathbf{e}_{js}
\exp (i \mathbf{k}_j \cdot \mathbf{r})
\end{equation}
defined in some large volume V (which may be
taken to be infinity later). For definiteness, we
shall work in the base vectors $\mathbf{e}_{js}$
corresponding to circular polarizations (and $s$
takes the values $\pm$). The annihilation and
creation operators $a_{js}$ and $a^\dagger_{js}$
obey the familiar commutation relations
\begin{equation}
[a_{js}, a^\dagger_{j^\prime s^\prime} ]
= \delta_{j j^\prime} \delta_{s s^\prime} .
\end{equation}

As pointed out by Karassiov~\cite{Kar93}, there
are specific observables that characterize proper
polarization properties of fields of the form
(\ref{Epol}). They correspond to the generators
of the group SU(2) of polarization gauge
invariance. In the circular polarization basis
they can be expressed as
\begin{eqnarray}
\label{polop}
& \displaystyle
J_+ = \sum_{j=1}^m a_{j +}^\dagger a_{j-} \, ,
\qquad
J_- = \sum_{j=1}^m a_{j -}^\dagger a_{j +} \, , &
\nonumber \\
& & \\
& \displaystyle
J_z = \frac{1}{2} \sum_{j=1}^m
( a_{j +}^\dagger a_{j +} - a_{j -}^\dagger a_{j -} ) \, .
&  \nonumber
\end{eqnarray}
They indeed furnish a Schwinger representation of the
su(2) algebra:
\begin{equation}
\label{su2ccr}
[J_z, J_\pm ] = \pm J_\pm \, , \qquad
[J_+, J_- ] = 2 J_z \, .
\end{equation}
The Casimir operator is
\begin{equation}
J^2 = J_z^2 + \frac{1}{2} ( J_+ J_- + J_- J_+ ) =
\frac{N}{2} \left ( \frac{N}{2} + \openone \right ) .
\end{equation}
The operator
\begin{equation}
\label{numop}
N = \sum_{j=1}^m
( a_{j +}^\dagger a_{j +} + a_{j -}^\dagger a_{j -} )
\end{equation}
represents the total number of photons and satisfies
\begin{equation}
[N, \mathbf{J}] = 0 \, ,
\end{equation}
where $\mathbf{J}= (J_x, J_y, J_z)$, with $J_\pm =
J_x \pm i J_y$. The total Hilbert space splits in
this way in invariant subspaces of dimension $N+1$.
Since there is no risk of confusion, we denote by
the same letter the operator (\ref{numop}) and
its eigenvalue, which is the total number of photons.

The quantities $\mathbf{J}$ are then measurable
in photon-counting experiments~\cite{Ray99}. In fact,
for a single-mode field they coincide up to a factor
1/2 with the Stokes operators~\cite{footnote}.
It is thus natural to extend this identification and
define
\begin{equation}
\mathbf{J} = \frac{1}{2} \mathbf{S}
\end{equation}
for $m$-mode fields. This allows one to
parallel the classical idea and define the
degree of polarization as
\begin{equation}
\label{Pcls}
P  = \frac{\sqrt{\langle \mathbf{S} \rangle^2}}
{\langle N \rangle} \, .
\end{equation}
Note that $\mathbf{s}= \langle \mathbf{S} \rangle /
\langle N \rangle $ is the polarization vector
in classical optics, which defines the Poincar\'e
sphere. Equation~(\ref{Pcls}) can then be
interpreted as the distance from the point
represented by $\mathbf{s}$ to the origin,
which is associated with the unpolarized
light.

To highlight the SU(2) invariance of
polarization, it has been proposed by several
authors~\cite{PC71,Aga71,LLP96,SBT01,Wun03} to
define unpolarized light as the field states that
remain invariant under any linear polarization
transformation,  which in experimental terms
can be accomplished with a combination of
phase plates and rotators (that produce
rotations of the electric field components
around the propagation axis). Any state satisfying
the invariance condition will also fulfill the
classical definition of an unpolarized state,
but the converse is not true.

The density operator of such quantum unpolarized
states can be written as
\begin{equation}
\varrho_{\mathrm{unpol}} = \bigoplus_{N}
r_N \ \openone_N \, ,
\end{equation}
where $\openone_N$ is the identity operator in
the invariant subspace with $N$ photons and all
$r_N$ are real and positive constants. The condition
of unit trace imposes
\begin{equation}
\sum_{N} ( N + 1 ) r_N = 1 .
\end{equation}
The vacuum state is the only pure quantum state that is
unpolarized, and the unpolarized mixed states are
totally mixed in each subspace. Note that in each
invariant subspace, $\varrho_{\mathrm{unpol}}$
is fully random: we claim that in any physical
depolarization process this fundamental structure
of invariant subspaces must be preserved.

\section{Light depolarization: single-mode fields}

\label{smf}

\subsection{Master equation for pure dephasing processes}

We begin by focusing our attention on a
single-mode field coupled to a bath system.
The total Hamiltonian is
\begin{equation}
H = H_{\mathrm{field}} +
H_{\mathrm{bath}} + H_{\mathrm{int}} \, .
\end{equation}
Here $H_{\mathrm{field}}$ is the Hamiltonian of
the mode under consideration
\begin{equation}
H_{\mathrm{field}} = \hbar \omega
\sum_{s= \pm} a_s^\dagger a_s \, ,
\end{equation}
while $H_{\mathrm{bath}}$ describes the free
evolution of the environment. We do not
make any hypothesis on the precise kind of
bath, and only assume that it is so large that
its statistical properties are unaffected by
the coupling with the system.

In the interaction picture and under the usual
weak-coupling and Markov approximations, the
master equation takes the general form~\cite{Gar00}
\begin{equation}
\label{tradme}
\dot{\varrho}(t) = - \frac{1}{\hbar^2}
\int_0^\infty d\tau \ \mathrm{Tr}_{\mathrm{B}}
\{ [H_{\mathrm{int}}(t), [H_{\mathrm{int}} (t - \tau ),
\varrho(t) \otimes \varrho_{\mathrm{B}} ]] \} \, ,
\end{equation}
where $\mathrm{Tr}_{\mathrm{B}}$ is the partial
trace over the bath variables and $\varrho_{\mathrm{tot}}
(t) = \varrho(t) \otimes \varrho_{\mathrm{B}}$
is the density operator for the total system.
In addition, we take the mode and the bath
initially independent. Henceforth, $\varrho (t) $
will denote the reduced density operator
for the field.

A common way to couple the system to the bath
is through the basic interaction Hamiltonian
\begin{equation}
\label{Com}
H_{\mathrm{int}} = \hbar
\sum_\lambda \sum_{s= \pm}
( \kappa_\lambda \ \Gamma_\lambda a_s^\dagger +
\kappa_\lambda^\ast \ \Gamma_\lambda^\dagger a_s ) \, ,
\end{equation}
where $\kappa_\lambda$ are coupling constants,
$\Gamma_\lambda$ and $\Gamma_\lambda^\dagger$ are
annihilation and creation operators for bath
quanta and the sum over $\lambda $ runs over
all the accessible bath modes. Note that a photon
can lose energy by creating a bath quantum, and
conversely.

We assume for simplicity that the bath is at
zero temperature, which is tantamount to
neglecting stimulated processes. The master
equation (\ref{tradme}) is then
\begin{equation}
\label{Linind}
\dot{\varrho} = \sum_{s} \frac{\gamma_s}{2}
\mathcal{L}[a_s] \ \varrho \, ,
\end{equation}
where $\mathcal{L} [ C_s ]$ are the Lindblad
superoperators~\cite{Lin76}
\begin{equation}
\mathcal{L}[C_s ] \ \varrho =
2 C_s \varrho C_s^\dagger -
\{ C_s^\dagger C_s, \varrho\} \, .
\end{equation}
and $\gamma_s$ are the decoherence rates associated
with each polarization. As was recognized
earlier~\cite{Kra71,GKS76}, this form of the master equation
is completely positive, which is the only dynamical
evolution of an open system ensuring that the state
of the system remains physically valid at all times.

Equation~(\ref{Linind}) corresponds to two
standard independent Lindblad decaying processes,
one for each basic polarization. Since the
coupling (\ref{Com}) transfers energy between
the system and the bath, the net result is a
damping of the beam intensity. In fact, one
can check that, irrespective of the initial
conditions, the stationary state is always
$\varrho (t \rightarrow \infty ) =
| 0,0 \rangle \langle 0,0 |$, where $| 0, 0
\rangle = | 0 \rangle_+ | 0 \rangle_-$ is the
vacuum state for both circular polarizations.
Obviously, this cannot describe depolarizing
processes.

As we have anticipated in the Introduction,
depolarization implies a pure dephasing process.
In mathematical terms, this means that the
interaction must commute with the system
Hamiltonian, so no energy is transferred
and only the phase changes. Various theoretical
scenarios have been proposed to that end.
Apart from minor details, all of them can be
modeled by an interaction Hamiltonian of the
type~\cite{Gar00}
\begin{equation}
\label{Gar}
H_{\mathrm{int}} = \sum_\lambda \sum_{s=\pm}
( \kappa_\lambda \ \Gamma_\lambda
a_s^\dagger a_s + \kappa_\lambda^\ast \
\Gamma_\lambda^\dagger a_s^\dagger a_s ) \, ,
\end{equation}
which can be viewed as a scattering process in
which a bath quantum can be absorbed or emitted,
but the number of photons in each polarization is
preserved.

Assuming again a zero-temperature bath, the master
equation for (\ref{Gar}) can be written as
\begin{equation}
\label{meGar}
\dot{\varrho} = \sum_{s=\pm}
\frac{\gamma_s}{2}
\mathcal{L} [ a_s^\dagger a_s ] \ \varrho \, .
\end{equation}
To illustrate the physics embodied in this equation,
let us focus on the fully quantum regime and
consider a one-photon state. We are then within
a two-dimensional invariant subspace, whose
basis will be labeled
\begin{equation}
\label{bas2d}
| + \rangle = | 1 \rangle_+ | 0 \rangle_- \, ,
\qquad
| - \rangle = | 0 \rangle_+ | 1 \rangle_- \, .
\end{equation}
The density matrix in this two-dimensional
subspace can be expressed as
\begin{equation}
\varrho = \frac{1}{2} \left (
\begin{array}{cc}
1+ s_z & s_x - i s_y \\
s_x + i s_y & 1- s_z
\end{array} \right ) =
\frac{1}{2} ( \openone +
\mathbf{s} \cdot \bm{\sigma} ) \, ,
\end{equation}
where $\mathbf{s}$ is the polarization vector
defined after Eq.~(\ref{Pcls}) and can be written
as $\mathbf{s} = 2 \Tr ( \varrho \bm{\sigma} )$,
$\bm{\sigma}$ being the Pauli matrices.
In this subspace the model (\ref{meGar}) can be
easily recast in terms of $\mathbf{s}$; the
final solution reads as
\begin{eqnarray}
\label{sol1}
s_{x}(t) & = & s_{x}(0)
e^{-(\gamma_+ + \gamma_- )t/2} \, ,  \nonumber \\
s_{y}(t) & = & s_{y}(0)
e^{-(\gamma _{+} + \gamma _{-})t/2}\,, \\
s_{z}(t) &=&s_{z}(0) \,.  \nonumber
\end{eqnarray}
This interaction indeed preserves the
invariant subspaces, but produces no
thermalization: its effect is merely to
maintain the occupation probabilities, while
erasing all coherences. Although this can
model appropriately many interesting phenomena,
it fails to describe depolarizing effects,
because depolarization leads to a uniform
distribution of photons in each invariant
subspace and not only to a pure dephasing.
In other words, depolarization not only
preserves the invariant subspaces, but
the steady state in each one of then must
be a diagonal state. We believe that these
conditions are essential to ensure a correct
description of depolarization.

\subsection{Modeling depolarization with a
nonresonant randomly-distributed atomic medium}

Apart from the previous drawbacks, generic
models of the type (\ref{Gar}) suffer from
the problem that one cannot provide a
physically feasible mechanism leading to
an interaction of that form. To solve these
difficulties we shall take another route.
We recall that in classical optics, the electric
field of the light experiences a decorrelation as
it propagates through a material medium. If
$\alpha_\pm $ denote the classical field-mode
amplitudes, such a decorrelation can be taken
into account by assuming that $\alpha_\pm \rightarrow
\alpha_\pm e^{i \varphi_\pm} $, where $\varphi_\pm$
are random phases.

We wish to explore this picture in the quantum world.
In consequence, we assume that the field propagates
through a material medium represented by a
collection of two-level atoms. Since there is
no net energy transfer between the field and
the atoms, their interaction must be necessarily
dispersive. In short, our basic system is
\begin{equation}
\label{Hsysdis}
H_{\mathrm{sys}} = H_{\mathrm{field}}+
H_{\mathrm{at}} + V \, ,
\end{equation}
where
\begin{eqnarray}
\label{btla}
H_{\mathrm{at}} & = & \sum_\lambda \frac{1}{2}
\hbar \omega_\lambda \sigma_\lambda^z \, ,
\nonumber \\
& & \\
V & = & \hbar \sum_\lambda \sum_{s= \pm}
( g_{\lambda s} \ \sigma_\lambda^-
a_s^\dagger + g_{\lambda s}^\ast \
\sigma_\lambda^+ a_s ) \, .  \nonumber
\end{eqnarray}
The form of the interaction $V$ assumes,
as it is usually done, that the atoms interact
with the field mode in the dipolar and rotating-wave
approximations. In addition, the atoms are  randomly
distributed so the coupling constants $g_{\lambda s}$
have random phases~\cite{Ors00}.

If $\Delta_\lambda = \omega_\lambda -\omega $ is
the detuning, we must consider the far off-resonant
regime $|g_{\lambda s}| \ll \Delta _{\lambda }$.
In such a limit, we can adiabatically eliminate
the nonresonant interactions in Eq.~(\ref{btla}) and
obtain the following effective Hamiltonian that
describes our system~\cite{Kli02} :
\begin{eqnarray}
H_{\mathrm{sys}} & \simeq &
\hbar \omega \mathcal{N} +
\sum_{\lambda }\frac{1}{2}\hbar
\Omega_\lambda  \sigma_\lambda^z
\nonumber  \label{Heff} \\
& + & \sum_{\lambda ,\lambda^\prime \atop
{\lambda \neq \lambda^\prime }}
\sum_{s = \pm } \frac{1}{2} \hbar
g_{\lambda s} g_{\lambda^\prime s}^\ast \
\left( \frac{1}{\Delta_\lambda} +
\frac{1}{\Delta_{\lambda^\prime}} \right)
\sigma_\lambda^+ \sigma_{\lambda^\prime}^-
\, ,  \nonumber \\
& &
\end{eqnarray}
where
\begin{equation}
\mathcal{N}= \sum_{s=\pm} a_s^\dagger a_s +
\sum_\lambda \frac{1}{2} \sigma_\lambda^z
\end{equation}
is the conserved excitation number operator and
\begin{eqnarray}
\label{defoJ}
\Omega_\lambda & = & \Delta_\lambda  +
\delta_\lambda + \frac{\mathcal{J}_\lambda}
{\Delta_\lambda} \, ,  \nonumber  \\
&& \\
\mathcal{J}_\lambda & = & 2 |g_\lambda|^2
( N + J_+ e^{i \varphi_\lambda} +
J_- e^{-i\varphi_\lambda} ) \, .
\nonumber
\end{eqnarray}
Here we have written $g_{\lambda \pm}  =
|g_\lambda | e^{ \pm i \varphi_\lambda/2}$,
with $\varphi_\lambda$ being random phases.
This reflects the fact that the relative
phase between the atomic dipole and the
field mode is random. The parameter
\begin{equation}
\delta_\lambda  = \frac{|g_\lambda|^2}
{\Delta_\lambda}
\end{equation}
represents just a small frequency shift.

Moreover, it is well known that the atoms
decay irreversibly. This usually assigned
to their interaction with the continuum of
modes of an additional thermal electromagnetic
environment. In such a case, the density matrix
for the system (\ref{Hsysdis}) evolves according
to~\cite{Gar00}
\begin{eqnarray}
\label{ime}
\dot{\varrho}_{\mathrm{sys}}(t) & = & -\frac{i}{\hbar}
[H_{\mathrm{sys}}, \varrho_{\mathrm{sys}}]  \nonumber \\
& + & \sum_\lambda \frac{\gamma_\lambda}{2}
\{(\bar{n}_\lambda + 1) \mathcal{L}
[\sigma_\lambda^- ] \ \varrho_{\mathrm{sys}} +
\bar{n}_\lambda \mathcal{L} [\sigma_\lambda^+ ]
\ \varrho_\mathrm{sys} \} \, , \nonumber \\
&&
\end{eqnarray}
where $\gamma_\lambda$ is the decay constant of
the $\lambda$th atom due to its coupling to the
thermal environment with $\bar{n}_\lambda$ excitations.
As  is well established~\cite{BP02,Kof01,JS04},
the properties of a random medium are well reproduced
in the high-temperature limit: $\bar{n}_\lambda
\gg 1$. In this limit, the effect of spontaneous emission
can be disregarded in comparison with the stimulated
emission processes. Emission into the reservoir and
absorption from the reservoir therefore become identical;
i.e., they balance each other in the stationary state:
the emission and absorption processes depend solely on
the initial population of the atomic state. Consequently,
the steady-state reduced density operator is
approximately given by a mixture of equally populated
atomic states. The density matrix of the $\lambda $th
atom thus becomes diagonal ($\varrho_\lambda  =
\frac{1}{2} \openone$) and the effect of the last
term in Eq.~(\ref{Heff}) is negligible and we shall
omit it henceforth. The effective Hamiltonian for
our system then reads as
\begin{equation}
\label{Heffta}
H_{\mathrm{sys}} \simeq
\hbar \omega \mathcal{N} +
\sum_{\lambda }\frac{1}{2}\hbar
\Omega_\lambda  \sigma_\lambda^z .
\end{equation}

As indicated in the Appendix, in this
limit we can also adiabatically eliminate
the atomic variables and obtain a master
equation that, after averaging over the
random phases, reads as
\begin{equation}
\label{mesb}
\dot{\varrho} = - i [\omega N, \varrho ]
+ \frac{\gamma}{2} \mathcal{L}[ N ] \ \varrho
+ \frac{\gamma}{2} \mathcal{L}[ J_+ ]\ \varrho
+ \frac{\gamma}{2} \mathcal{L}[ J_- ] \ \varrho \, ,
\end{equation}
where $\varrho (t) = \Tr_{\mathrm{at}}[
\varrho_{\mathrm{sys}}(t)]$ is again the
reduced density operator for the field mode
and the decoherence rate $\gamma $ is
\begin{equation}
\label{gam}
\gamma = 4 \sum_\lambda \frac{|g_\lambda|^4}
{\gamma_\lambda \Delta_\lambda^2 \bar{n}_\lambda} \,.
\end{equation}
Equation (\ref{mesb}) is our central result. We
observe that it  preserves the SU(2) invariant
subspaces and the steady state in each $N$-photon
subspaces is a completely random state
\begin{equation}
\label{ss}
\varrho ( t \rightarrow \infty ) =
\frac{1}{N+1} \openone \, .
\end{equation}
This can be considered as the major advantage
of our approach.

A rough estimate immediately shows that the
depolarization rates (\ref{gam}) appearing in
this model are very small when compared with
other typical system parameters, which is in
agreement with the experimental observations.

The terms $\mathcal{L} [ J_{\pm} ] \ \varrho$
describe depolarization in each invariant subspace,
meanwhile the action of $\mathcal{L} [ N ] \
\varrho $ therein is trivial. Nevertheless, this
$\mathcal{L} [ N ] \ \varrho $ is responsible
for the relative phase decay between blocks
of the density matrix corresponding to
different excitation numbers.

To compare with the previous discussion
in Eq.~(\ref{sol1}), we consider again the
one-photon case. By recasting the depolarization
master equation (\ref{mesb}) in terms of $\mathbf{s}$,
one obtains the solution as
\begin{eqnarray}
s_x (t) & = & s_x (0) e^{- \gamma t} \, ,
\nonumber \\
s_y (t) & = & s_y (0) e^{- \gamma t} \, , \\
s_z (t) & = & s_z (0) e^{-2 \gamma t} \, ,
\nonumber
\end{eqnarray}
which displays the aforementioned desirable properties.
The degree of polarization (\ref{Pcls}) of this mode
evolves then as
\begin{equation}
P (t) = [s_x^2 ( 0 ) + s_y^2 ( 0 ) + s_z^2 ( 0 )
e^{- 2 \gamma t}]^{1/2} e^{-\gamma t} \, .
\end{equation}

\section{Light depolarization: multimode fields}

\label{mmf}

For the multimode case, it seems reasonable to
assume that the depolarization is independent
in each mode. The corresponding master equation
can be thus simply written as
\begin{eqnarray}
\label{meguay2}
\dot{\varrho} & =  & -i \sum_{j=1}^m
\omega_j [ N_j, \varrho ] \nonumber \\
& + & \sum_{j=1}^m \frac{\gamma_j}{2}
\{ \mathcal{L}[ N_j] \ \varrho +
\mathcal{L}[J_{j +}] \ \varrho +
\mathcal{L}[J_{j -}] \ \varrho \} \, .
\end{eqnarray}
One can check that the properties discussed in the
previous section also hold in this multimode
case.

For concreteness, henceforth we restrict our
attention to the case of a two-mode field. We
have then a representation of SU(2) $\otimes $
SU(2), and the whole Hilbert space can be
decomposed in irreducible subspaces. For the
case where there is one photon in each mode,
we have, using the standard terminology of
SU(2) representations, that $D^{1/2} \otimes D^{1/2}
= D^0 \oplus D^1$, where $D^j$ is the invariant
subspace with eigenvalue $j$.

The one-dimensional subspace $D^0$ is spanned by
the singlet state
\begin{equation}
| \psi^{-} \rangle = \frac{1}{\sqrt{2}}
( | + \rangle_1 \, | - \rangle_2 \, -
\, | - \rangle_1 \, | + \rangle_2 ) \, ,
\end{equation}
while $D^1$ is spanned by the triplet states
\begin{equation}
| \psi^+ \rangle =
\frac{1}{\sqrt{2}} ( | + \rangle_1 \, | -
\rangle_2 \, + \, | - \rangle_1 \, | + \rangle_2 ) ,
\;
| + \rangle_1 \, | + \rangle_2 ,
\;
| - \rangle_1 \, | - \rangle_2
\, .
\end{equation}
Here the states $| \pm \rangle$ for each mode are
given as in (\ref{bas2d}) and the subscripts 1 and
2 label the corresponding modes.

We take our system to be initially in a generic
density matrix such as
\begin{equation}
\varrho (0) = \left (
\begin{array}{llll}
\varrho_{11} (0) & \varrho_{12} (0) &
\varrho_{13} (0) & \varrho_{14} (0) \\
\varrho_{21} (0) & \varrho_{22} (0) &
\varrho_{23} (0) & \varrho_{24} (0) \\
\varrho_{31} (0) & \varrho_{32} (0) &
\varrho_{33} (0) & \varrho_{34} (0) \\
\varrho_{41} (0) & \varrho_{42} (0) &
\varrho_{43} (0) & \varrho_{44} (0)
\end{array}
\right ) \, ,
\end{equation}
where we have employed the standard
eigenbasis
\begin{equation}
\label{sta}
\begin{array}{ll}
|1 \rangle = | + \rangle_1 \, | + \rangle_2
\, , & \quad
|2 \rangle = | + \rangle_1 \, | - \rangle_2
\, , \\
&  \\
|3 \rangle = | - \rangle_1 \, | + \rangle_2
\, , & \quad
| 4 \rangle = |- \rangle_1 \, | - \rangle_2 \, .
\end{array}
\end{equation}
After some computation, the solution of the
master equation (\ref{meguay2}) can be shown
to be
\begin{eqnarray}
\label{cogcoh}
\varrho_{21}(t) & = & \frac{1}{2}
[\varrho _{21}(0)(1+e^{-2\gamma_1t})
\nonumber \\
& + & \varrho_{43}(0) (1 -
e^{-2\gamma_1 t})]e^{-\gamma_2 t}
\, ,  \nonumber \\
\varrho_{31}(t) & = & \frac{1}{2}
[\varrho _{31}(0)(1+e^{-2\gamma_2 t})
\nonumber \\
& + & \varrho _{42}(0) (1 -
e^{-2\gamma_2 t})]e^{-\gamma_1 t}
\, ,  \nonumber \\
\varrho_{41} (t) & = &
\varrho _{41}(0) \,
e^{-(\gamma_1+\gamma_2)t}
\, , \nonumber \\
\varrho_{32}(t) & = & \varrho _{32}(0) \,
e^{-(\gamma_1+\gamma_2)t}
\, , \nonumber \\
\varrho_{42} (t) & = & \frac{1}{2}
[\varrho_{42}(0)( 1 + e^{-2\gamma_2 t})
\nonumber \\
& + & \varrho_{31}(0)( 1 -
e^{-2\gamma_2 t})]e^{-\gamma_1 t}
\,,  \nonumber \\
\varrho_{43}(t) & = & \frac{1}{2}
[\varrho_{43}(0)( 1 + e^{-2\gamma_1 t})
\nonumber \\
& + & \varrho_{21}(0) ( 1 -
e^{-2\gamma_1 t})]e^{-\gamma_2 t} \,,
\end{eqnarray}
for the nondiagonal elements, and
\begin{eqnarray}
\label{cogdia}
\varrho_{11}(t) & = & \frac{1}{4}
(1 + \{ 2 [\varrho_{11}(0) + \varrho_{22}(0)]- 1 \}
e^{-2\gamma_1 t}  \nonumber \\
& + & \{2 [ \varrho_{11}(0) + \varrho_{33}(0)]-1 \}
e^{-2\gamma_2 t}  \nonumber \\
& + & \{ 2 [\varrho_{11}(0) + \varrho_{44}(0)]-1 \}
e^{-2(\gamma_1 + \gamma_2)t}) \, ,  \nonumber \\
\varrho_{22} (t) & = & \frac{1}{4}
(1 + \{ 2 [\varrho_{22}(0) + \varrho_{11}(0) ]-1 \}
e^{-2\gamma_1 t}  \nonumber \\
& + & \{ 2 [\varrho_{22} (0) + \varrho_{44}(0)]-1 \}
e^{-2\gamma_2 t}  \nonumber \\
& + & \{2 [\varrho_{22} (0) + \varrho_{33}(0)]-1 \}
e^{-2(\gamma_1 + \gamma_2)t}) \,,  \nonumber \\
\varrho_{33} (t) & = & \frac{1}{4}
( 1 + \{ 2 [\varrho_{33} (0) + \varrho_{44}(0) ]- 1 \}
e^{-2\gamma_1 t}  \nonumber \\
& + & \{ 2 [\varrho_{33}(0) + \varrho_{11}(0)]- 1 \}
e^{-2\gamma_2 t}  \nonumber \\
& + & \{ 2 [\varrho_{33} (0) + \varrho_{22} (0)]-1 \}
e^{-2(\gamma_1 + \gamma_2 )t}) \,,
\end{eqnarray}
for the diagonal ones. Here $\gamma_1$ and
$\gamma_2$ are the decoherence rates for modes 1 and
2, respectively. Note that the trace condition
gives $\varrho_{44} = 1 - \varrho_{11} - \varrho_{22} -
\varrho_{33}$.

In this basis, the operators $\mathbf{J}$ defined in
Eq.~(\ref{polop}), take the explicit form
\begin{eqnarray}
&
J_x = \frac{1}{2}
\left (
\begin{array}{rrrr}
0 & 1 & 1 & 0 \\
1 & 0 & 0 & 1 \\
1 & 0 & 0 & 1 \\
0 & 1 & 1 & 0
\end{array}
\right ) \, ,
\qquad
J_y = \frac{1}{2}
\left (
\begin{array}{rrrr}
0 & -i & -i & 0 \\
i & 0 & 0 & -i \\
i & 0 & 0 & -i \\
0 & i & i & 0
\end{array}
\right ) \, ,
& \nonumber \\
& & \\
&
J_z =
\left (
\begin{array}{rrrr}
1 & 0 & 0 & 0 \\
0 & 0 & 0 & 0 \\
0 & 0 & 0 & 0 \\
0 & 0 & 0 & -1
\end{array}
\right ) \, .
& \nonumber
\end{eqnarray}
In consequence, the degree of polarization (\ref{Pcls})
is now
\begin{eqnarray}
P (t) & = & [ | \varrho_{12} (t) + \varrho_{13} (t) +
\varrho_{24} (t) + \varrho_{34} (t) |^2 \nonumber \\
& + & | \varrho_{11} (t) - \varrho_{44} (t)|^2 ] \, .
\end{eqnarray}
This is a quite compact result. Consider, for example,
a pure disentangled state such as any element of the
basis (\ref{sta}), say $\varrho (0) = | 1 \rangle
\langle 1 |$. Using (\ref{cogcoh}) and (\ref{cogdia})
we immediately get
\begin{equation}
P (t) = \frac{1}{2} (e^{-2 \gamma_1 t} +
e^{-2 \gamma_2 t}) \, .
\end{equation}
On the other hand, for the maximally entangled Bell-like
states~{\cite{Gal02}}
\begin{equation}
|\psi^{\pm } \rangle \langle \psi ^{\pm }| =
\frac{1}{2}
\left (
\begin{array}{rrrr}
0 & 0 & 0 & 0 \\
0 & 1 & \pm 1 & 0 \\
0 & \pm 1 & 1 & 0 \\
0 & 0 & 0 & 0
\end{array}
\right) \, ,
\end{equation}
we have
\begin{equation}
P (t) = \frac{1}{2}  e^{- ( \gamma_1 + \gamma_2) t} \, .
\end{equation}
Now, owing to the strong correlations, the decoherence
rates for both modes always appear together. We stress
that even the singlet state will become depolarized. This
is due to the fact that the depolarization mixes
different invariant subspaces. Note also that although
these two examples are pure states, they show  quite a
different evolution of the degree of polarization.

\section{Concluding remarks}
\label{con}

What we expect to have accomplished in this
paper is to present a comprehensive theory of
the depolarization of quantum fields. In our
model the field modes couple dispersively
to a randomly distributed atomic bath:
the resulting master equation has unique
properties that we have explored in detail.

We hope that the tools introduced here could
be of interest in treating depolarization in
the fully quantum regime, especially for
biphotons, which have strong implications in
areas of futuristic technologies such as quantum
computing, quantum cryptography, and quantum
communications.

\appendix

\section{Derivation of the master equation (\ref{mesb})}

Let us define the following operators
\begin{equation}
\begin{array}{ll}
\label{so}
N_\lambda^+ \varrho = \sigma_\lambda^+
\varrho \sigma_\lambda^- \, ,
\qquad  &
N_\lambda^- \varrho = \sigma_\lambda^-
\varrho \sigma_\lambda^+ \, , \\
&  \\
\displaystyle
N_\lambda^l \varrho = \frac{1}{2}
\sigma_\lambda^z \varrho \, ,
\qquad  &
N_\lambda^r \varrho = \frac{1}{2}
\varrho \sigma_\lambda^z \, , \\
&  \\
A_\lambda^l \varrho =
\Omega_\lambda \varrho \, ,
\qquad  &
A_\lambda^r \varrho = \varrho \Omega_\lambda \, ,
\nonumber
\end{array}
\end{equation}
which satisfy the following commutation relations
\begin{eqnarray}
\ [ N_\lambda^\pm , N_{\lambda^\prime}^l]  & = &
\mp N_\lambda^\pm \delta_{\lambda \lambda^\prime}
\, ,  \nonumber \\
\ [ N_\lambda^\pm , N_{\lambda^\prime}^r] & = &
\mp N_\lambda^\pm \delta_{\lambda \lambda^\prime}
\, , \\
\ [ N_\lambda^\alpha ,A_\lambda^\beta ] & = & 0 \,,
\nonumber
\end{eqnarray}
where $\alpha ,\beta =\pm ,l,r$.
Apart from their interesting algebraic
properties, about which we are not concerned in
this paper, the operators (\ref{so}) allow one
to recast the master equation~(\ref{ime}) in a more
appropriate form. In fact in the high-temperature limit
we can replace $H_{\mathrm{sys}}$ by the effective
Hamiltonian (\ref{Heffta}) and we have
\begin{eqnarray}
\label{rhoA1}
\dot{\varrho}_{\mathrm{sys}} & = & -
\sum_\lambda \gamma_\lambda \bar{n}_\lambda
\varrho_{\mathrm{sys}}   \nonumber \\
& + & \sum_\lambda [ \gamma_\lambda \bar{n}_\lambda
(N_\lambda^+ + N_\lambda^-) -
2 i A_\lambda^l N_\lambda^r +
2 i A_\lambda^r N_\lambda^r ] \varrho_\mathrm{sys} \, .
\nonumber \\
\end{eqnarray}
The dissipative term in this equation is much bigger
than the Hamiltonian term. This allows us to adiabatically
eliminate the atomic degrees of freedom. We can e.g.
make use of the perturbative method proposed in
Refs.~\cite{Kli00} and \cite{SD01} (see  Ref.~\cite{KRD03}
for a complete account of the application of the
method to effective master equations). This requires
us to represent Eq.~(\ref{rhoA1}) in terms of diagonal
and raising-lowering operators. To this end, we first
diagonalize the dissipative part of (\ref{rhoA1}) by
applying the following $\pi /2$ rotation
\begin{equation}
\label{V}
V = \exp \left [ \frac{\pi }{2} \sum_\lambda
\frac{1}{2}( N_\lambda^+ - N_\lambda^- ) \right] \,,
\end{equation}
which leads to
\begin{eqnarray}
\dot{\varrho}_{\mathrm{sys}} & = &
-\sum_\lambda [ \gamma_\lambda \bar{n}_\lambda
+ \gamma_\lambda \bar{n}_\lambda
( N_\lambda^r + N_\lambda^l )/2
\nonumber \\
& - & i ( A_\lambda^r - A_\lambda^l )
( N_\lambda^+ + N_\lambda^- ) \nonumber \\
& + & i ( A_\lambda^r + A_\lambda^l )
( N_\lambda^l - N_\lambda^r ) ]
\varrho_{\mathrm{sys}} \, .
\end{eqnarray}
Next, we apply the following ``small rotation"
\begin{equation}
U =  \exp \left[ -i
\sum_\lambda \frac{A_\lambda^r - A_\lambda^l}
{2 \gamma_\lambda \bar{n}_\lambda}
( N_\lambda^+ - N_\lambda^-) \right] \, .
\end{equation}
After a lengthy but otherwise straightforward
calculation and applying the inverse of  the
transformation (\ref{V}) one finally gets
\begin{eqnarray}
 \label{A8}
 \dot{\varrho}_{\mathrm{sys}} & = & -
\sum_\lambda \{ \gamma_\lambda
\bar{n}_\lambda [ 1 - N_\lambda^+ - N_\lambda^- ]
\nonumber \\
& + & i( A_\lambda^r + A_\lambda^l )
( N_\lambda^l - N_\lambda^r ) \} \varrho_{\mathrm{sys}}
\nonumber \\
& - & \sum_\lambda 2 \gamma_\lambda \bar{n}_\lambda
\left( \frac{A_\lambda^r - A_\lambda^l}
{2 \gamma_\lambda \bar{n}_\lambda} \right)^2
( N_\lambda^+ + N_\lambda^- ) \varrho_{\mathrm{sys}} \,,
\end{eqnarray}
where the field mode is expressed in a rotating frame.
This transformed equation has the virtue of containing
only diagonal terms so we can immediately trace over
atomic variables to obtain the master equation for the
density matrix of the mode  $\varrho (t) = \Tr_{\mathrm{at}}
[\varrho_{\mathrm{sys}} (t)]$. Since each atom in the
medium is in a statistical mixture, $\varrho_\lambda =
\frac{1}{2} \openone$, the first term in (\ref{A8})
vanishes and one obtains
\begin{equation}
\dot{\varrho} = \sum_\lambda \frac{1}{2 \gamma_\lambda
\Delta_\lambda \bar{n}_\lambda}
\mathcal{L}[ \mathcal{J}_\lambda] \ \varrho \, ,
\end{equation}
where $\mathcal{J}_\lambda$ has been defined in
Eq.~(\ref{defoJ}). If we average over all the
random phases $\varphi _{\lambda }$ we get
\begin{equation}
\dot{\varrho} = \frac{\gamma}{2}
\mathcal{L}[N]\ \varrho +
\frac{\gamma}{2} \mathcal{L}[ J_+ ] \ \varrho +
\frac{\gamma}{2} \mathcal{L}[ J_- ] \ \varrho \, ,
\end{equation}
with $\gamma $ given in (\ref{gam}). This is precisely
the master equation (\ref{mesb}), written in the
interaction picture.

\begin{acknowledgments}
In the long process of refining this manuscript we
have benefited from discussions with many colleagues.
We wish to thank especially A. P. Alodjants, G. M.
D' Ariano, G. Bj\"{o}rk, A. Galindo, H. de Guise,
M. Raymer, J. S\"{o}derholm, and A. I. Solomon.
\end{acknowledgments}


\begin{thebibliography}{99}
\bibitem{footnote}
To avoid any misunderstanding, we note that what
we are calling here single-mode fields, are also
known as two-mode fields, especially when dealing
with polarization properties.
\end{thebibliography}
\end{document}